\documentstyle[prl,aps,epsfig]{revtex}

\newcommand{\epem}{e^{+}e^{-}}
\newcommand{\ee}{e^{+}e^{-}}
\newcommand{\leplep}{\ell^{+}\ell^{-}}
\newcommand{\jp}{J/\psi}
\newcommand{\mumu}{\mu^{+}\mu^{-}}
\newcommand{\pipi}{\pi^{+}\pi^{-}}
\newcommand{\ks}{K^{0}}

\newcommand{\rt}{\rightarrow}

\newcommand{\etal}{\em et al.}

\begin{document}
\draft
\title{Observation of $B\rt\jp K_1(1270)$}
\author{The Belle Collaboration}

\author{
  K.~Abe$^{10}$,              
  K.~Abe$^{38}$,              
  I.~Adachi$^{10}$,           
  Byoung~Sup~Ahn$^{16}$,      
  H.~Aihara$^{40}$,           
  M.~Asai$^{11}$,             
  Y.~Asano$^{45}$,            
  T.~Aso$^{44}$,              
  V.~Aulchenko$^{2}$,         
  T.~Aushev$^{14}$
% ...
}

\address{
% ...
$^{2}${Budker Institute of Nuclear Physics, Novosibirsk}\\
% ...
$^{10}${High Energy Accelerator Research Organization (KEK), Tsukuba}\\
$^{11}${Hiroshima Institute of Technology, Hiroshima}\\
% ...
$^{14}${Institute for Theoretical and Experimental Physics, Moscow}\\
% ...
$^{16}${Korea University, Seoul}\\
% ...
$^{38}${Tohoku Gakuin University, Tagajo}\\
% ...
$^{40}${University of Tokyo, Tokyo}\\
% ...
$^{44}${Toyama National College of Maritime Technology, Toyama}\\
$^{45}${University of Tsukuba, Tsukuba}\\
% ...
}

\date{\today}
\maketitle
\begin{abstract}

We report the first observation of the exclusive
decay process $B\rt\jp K_1(1270)$ using a sample
of 11.2M $B\overline{B}$ meson pairs collected in
the Belle detector at the KEKB asymmetric energy $e^+e^-$
collider.  We measure 
branching fractions of
${\cal B}(B^0\rt \jp K^0_1(1270))
= (1.30 \pm 0.34 \pm  0.31)\times 10^{-3}$ and
${\cal B}(B^+\rt \jp K^+_1(1270))
=  (1.80 \pm 0.34 \pm  0.39)\times 10^{-3} $,
where the first error is statistical and the second is systematic.
These modes constitute approximately 15\% of
the total number of $B\rt\jp X$ decays.  
No evidence is seen for  $B\rt\jp K_1(1400)$ and we set
an upper limit for this branching fraction. 
The $K_1(1270)\rt \ks\pipi$ decays have a substantial $\ks\rho^0$ 
intermediate state component that may be useful for 
$CP$ violation studies.
\end{abstract}

\narrowtext
\twocolumn

Decays of $B$ mesons into final states containing 
the $\jp$ charmonium state 
play a special role in studies of $CP$ violation physics. 
Since the $\jp$ is itself a $CP$ eigenstate,
final states where the accompanying particles are matter-antimatter
symmetric are potentially useful for $CP$ violation measurements.
Moreover, these decay modes are experimentally convenient, primarily 
because the $\jp\rt\leplep$ ($\leplep = \epem~{\rm or}~\mumu$) final
states have a rather distinct signature.

However, although the branching fraction for inclusive $B\rt\jp X$ decay
is relatively large ($\sim 1\% $), only a small fraction
of these decays have been associated with exclusive decay
modes that are relevant for $CP$ studies.  Since
all current experimental searches for $CP$ violations in $B$ meson
decays are statistics limited,
it is important to identify additional
decay modes that might be useful. 
Decays of the type $B^0\rt\jp K^0_1(1270)$ are of
interest because the $K^0_1(1270)$ has an
appreciable branching fraction to the 
{\em flavor-nonspecific} $\ks\rho^0$ final state (14\%) and, thus,
might be useful for $CP$ measurements. 
At present there is very little experimental
information available about any
exclusive $B\rt\jp K\pi\pi$ decay modes~\cite{kpipi}.

In this Letter we describe a study of the $B\rt\jp K\pi\pi$ decay
process using the Belle detector~\cite{Belle} at the KEKB asymmetric energy
$\ee$ storage ring~\cite{KEKB}.  We
observe a signal for exclusive decays where
the properties of the $K\pi\pi$ system are consistent with those of the
$K_1(1270)$ resonance.  The data sample corresponds
to an integrated luminosity of  10.5~fb$^{-1}$
accumulated at the $\Upsilon(4S)$ resonance and contains
11.2M $B\overline{B}$ meson pairs.

The Belle detector consists of a three-layer silicon vertex detector,
a 50-layer central drift chamber (CDC) for charged particle
tracking and specific ionization measurements ($dE/dx$), an array of 1188
aerogel ${\rm \check{C}}$erenkov counters
(ACC), a time-of-flight (TOF) system comprised of 128 scintillation
counters,
and an electromagnetic calorimeter containing 8736 CsI(Tl) 
crystals (ECL), all located 
inside a 3.4 m diameter superconducting solenoid that generates
a 1.5~Tesla magnetic field.
An iron flux-return yoke outside the solenoid is
comprised of 14 layers of 4.7~cm-thick iron 
plates interleaved with a system of resistive plate counters (KLM)
that are used for muon identification and $K_L$ detection.  
Electron identification is based on a combination of
CDC $dE/dx$ information,
the response of the ACC, and the position, shape and energy 
deposit of the associated ECL shower.
Muon identification relies on the location and penetration
depth of associated tracks in the KLM.  We use a Monte Carlo
(MC) simulation to model the response of the detector
and determine acceptances~\cite{MC}.
The detector is described in detail in ref.~\cite{Belle}.

We select events with three final state topologies:
$B^+ \rt \jp K^+\pipi $,  and 
$B^0 \rt  \jp K^+\pi^- \pi^0 $ and $\jp \ks\pipi $,
where $\jp\rt\leplep$ and $\ks\rt\pipi$. 
We use $ B^+  \rt  \jp K^+$  decays for  normalization.
(Here, as in the rest of this report,
inclusion of the charge conjugate states is implied.)
Candidate $\jp\rt\mumu$ decays are oppositely charged
track pairs where at least one track is positively
identified as a muon  and the other
is either positively identified as a muon or has an associated ECL
energy deposit that is consistent with a minimum ionizing particle.
The invariant mass of the candidate $\mumu$ pair
is required to be within $\pm 3\sigma$ of the
$\jp$ mass peak, where $\sigma \simeq  12$~MeV is the mass resolution.
Candidate  $\jp\rt\ee$ decays are oppositely charged  track pairs
where at least one track is well identified as an electron
and the other track satisfies at least either the $dE/dx$ or the
ECL electron identification requirements.  
In this channel, we partially correct for final state
radiation or real bremsstrahlung in the inner parts of the detector
by including the four-momentum of every photon detected within 0.05~radians
of the original $e^+$ or $e^-$ direction in the $\ee$ invariant mass
calculation.   Since the $\jp\rt\ee$ peak still has
a residual radiative tail,
we use an asymmetric invariant mass requirement 
$-7\sigma \le (M_{\ee}-M_{\jp}) \le 3\sigma$ ($\sigma\simeq 12$~MeV).
After selection, the $J/\psi$ candidate tracks are refitted to
a common vertex and then mass constrained to  $M_{J/\psi}$.
The fitted vertex is then used as the vertex point 
of reference for the other particles in the decay.

For charged hadron identification we use the combination of
CDC $dE/dx$ measurements, flight times measured in the TOF, 
and the response of the ACC to determine a relative $K/\pi$ discrimination  
variable that ranges from ${\cal P}_{K/\pi} = 0$ 
for unambiguous pions, to ${\cal P}_{K/\pi} = 1$ 
for  well identified kaons.  In this analysis we identify
charged tracks with ${\cal P}_{K/\pi} > 0.5$
as kaons and those with ${\cal P}_{K/\pi} \le 0.5$ as pions.  
For $\pi^0\rt\gamma\gamma$ candidates, we require 
a minimum $\gamma$ energy of $40$~MeV  
and use $\gamma\gamma$ pairs with a total
laboratory energy greater than 230~MeV and
an invariant mass that is within $\pm 2\sigma$ of $M_{\pi^0}$, 
where the average value of $\sigma$ is $4.9$~MeV.
For $\ks\rt\pipi,$ we use oppositely charged track pairs where
the two-track vertex is displaced from that of the $\jp$ in the
transverse ($r$-$\phi$) plane by more than 0.8~mm, the 
$\phi$ directions of the vertex point and
the $\ks$ candidate's three momentum vector
agree within 0.2~radians, and the $\pipi$ invariant mass is within $\pm3\sigma$
of $M_{\ks}$, where $\sigma \simeq 4$~MeV.
We eliminate events of the type
$B^+\rt\psi(2S) K^+$, where $\psi(2S)\rt\pipi\jp$, by requiring
$M_{\pi\pi\ell\ell} - M_{\ell\ell}$ to differ from
$M_{\psi(2S)} - M_{\jp}$ by more than $\pm 9$~MeV
($3\sigma$).

Candidate $B$ mesons are identified by their 
center of mass (cm) energy difference, 
$\Delta E =  \sum_{i}{E_i} - E_{b}$, and the
beam constrained mass, 
$M_{bc}=\sqrt{E_{b}^2  - (\sum_{i}\vec{p}_i)^2}$, 
where $E_{b}=\sqrt{s}/2$ is the beam energy in the cm frame and 
$\vec{p}_i$ and $E_i$ are the cm three-momenta and energies
of the  candidate $B$ meson decay products.
We select  events with $M_{bc}>5.25$~GeV and $|\Delta E|<0.2$~GeV, and
define a $\pm 3\sigma$  {\em signal} window
of $|M_{bc}- M_{B}|< 9$~MeV and $|\Delta E |<60$~MeV
and an equal area {\em sideband} region that corresponds to the same
$M_{bc}$ selection and 
$60~{\rm MeV} < \Delta E  \le 180$~MeV~\cite{sideband}.

About 10\% of selected events have more than one entry in the
$M_{bc}>5.25$~GeV and $|\Delta E|<0.2$~GeV region.  
Multiple entries from additional charged pions are primarily
due to extra tracks produced by
low momentum curling particles ($p_t<0.25$~GeV/c);
fake $\pi^0$s are due to 
low energy $\gamma$ combinatoric background.
For multiple charged track entries,
we select the one with the
smallest impact parameter relative
to the refitted $\jp$ vertex.  According to the MC simulation,
these criteria select the correct track in 75\% of the cases.
For multiple entries involving $\pi^0$ candidates, 
we chose the $\gamma\gamma$
combination with an invariant mass closest to $M_{\pi^0}$.  This
selects the correct $\gamma\gamma$ combination in 60\% of the cases.

\begin{figure}[tb]
\begin{center}
\epsfxsize 3.3 truein \epsfbox{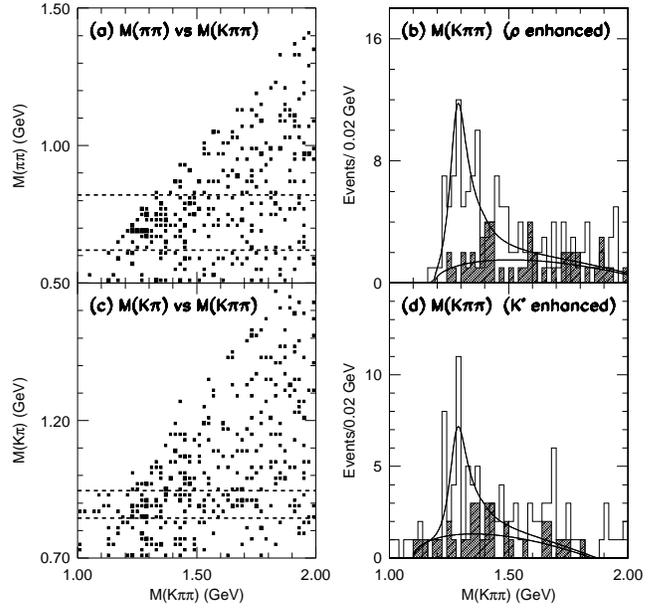}
\end{center}
\caption{{(a)} The distribution of $M_{\pipi}$ {\em vs}
$M_{K^+\pipi}$ for $B\rt \jp K\pi\pi$ candidates.
The dashed lines indicate the $\rho\rt\pi\pi$ selection region. 
{(b)} The  $K\rho$ mass distribution for the 
signal  (clear histogram) and sideband (cross-hatched histogram) regions.  
{ (c)} $M_{K\pi}$ {\em vs} $M_{K^+\pipi}$ for the same events.
The dashed lines indicate the $K^*\rt K\pi$ selection region.
{(d)} The $K^*\pi$ mass distribution for the 
signal  (clear histogram) and sideband (cross-hatched histogram) regions.  
The curves are the results of the fits described in the text.
}
\label{fig:Mpipi_vs_MKpipi}
\end{figure}

Figure~\ref{fig:Mpipi_vs_MKpipi}(a) shows the
distribution of $M_{\pipi}$ {\em vs}
$M_{K^+\pipi}$ for events in the $M_{bc}$ and $\Delta E$ signal window.
The clustering near $M_{\pi\pi}\simeq M_{\rho}$ and
$M_{K\pi\pi}\simeq 1.27$~GeV is consistent with expectations
for  $K_1(1270)\rt K\rho$ decays.  In these decays, the $K\pi\pi$
and $\pi\pi$ systems are
produced very near the kinematic boundary, which distorts the resonance
line shapes for both the $\rho$ and the $K_1(1270)$.
Figure~\ref{fig:Mpipi_vs_MKpipi}(b) shows the
$M_{K^+\pipi}$ distribution for the  
events with $-150~{\rm MeV}<(M_{\pi\pi} -M_{\rho})<50$~MeV~\cite{rho}.
In the figure, the clear histogram represents the events in the
$M_{bc}$ {\em vs.} $\Delta E$ signal window; sideband region events
are shown as the cross-hatched histogram. The lower curve shows the
result of a fit to a phase-space-like
background function to the sideband $M_{K\pi\pi}$ distribution.
The upper curve shows the results
of a fit to the mass region below 1.7~GeV
that uses the background
determined from the sideband
distribution plus
a line shape function that was specialized~\cite{WMD}
to expectations
for $K_1\rt K\rho$ decays with the PDG values for the 
$K_1$ mass and total width used as input~\cite{PDG}.
The sideband
background plus the $K_1(1270)$
line shape function, which has only its normalization as a free
parameter, gives a good fit to the lower part of the
$K\pi\pi$ mass spectrum~\cite{himass}, 
indicating that our interpretation of 
the event cluster in Fig.~\ref{fig:Mpipi_vs_MKpipi}(a) as being due
to the $K_1(1270)$ is reasonable.

In flavor-SU(3), the strange axial-vector eigenstates
are mixtures of the $K_1(1270)$ with the $K_1(1400)$, which
decays primarily via $K^*(890)\pi$ (Bf$\simeq 94\%$).
We searched the same events for a $K_1(1400)$ signal in
the distribution of $M_{K^+\pi^-}$ {\em vs}
$M_{K^+\pipi}$  
shown in Fig.~\ref{fig:Mpipi_vs_MKpipi}(c).
Here there is no obvious concentration of events around
$M_{K\pi}\simeq M_{K^*}$ and $M_{K\pi\pi}\simeq M_{K_1(1400)}$.
Figure~\ref{fig:Mpipi_vs_MKpipi}(d) shows the $M_{K\pi\pi}$ projection
for events with $\vert M_{K\pi} - M_{K^*}\vert < 50$~MeV,
where again there is no evidence for a $K_1(1400)$ signal; the
$M_{K\pi\pi} < 1.7$~GeV mass region is well fitted by
a sideband background plus the $K_1(1270)$.

We select $B\rt\jp K_1(1270)$ candidate events as those 
that satisfy the $\rho$ mass requirements and
have $1.16~{\rm GeV}\le M_{K\pi\pi}\le 1.38$~GeV.  The
MC simulation indicates that this mass window accepts
41\% of all $K_1(1270)\rt K\pi\pi$ decays.

\begin{figure}[tb]
\begin{center}
\epsfxsize 3.3 truein \epsfbox{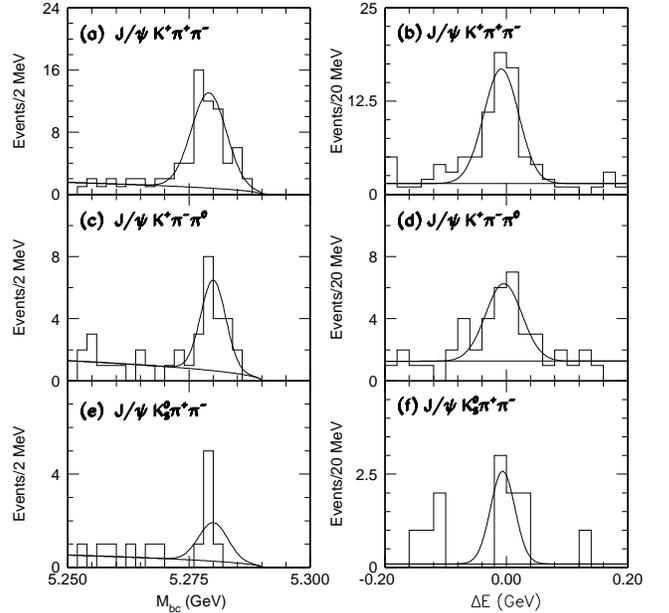}
\end{center}
\caption{ {(a)} The $M_{bc}$ and {(b)} $\Delta E$ projections
for the $K^+\pipi$ channel.  The fits are described in the text.
The corresponding distributions and fits for the
$K^+\pi^-\pi^0$ 
and $K^0_S\pipi$ channels are also shown.}
\label{fig:de_mb_fits}
\end{figure}

Figures ~\ref{fig:de_mb_fits}(a) through
~\ref{fig:de_mb_fits}(f)  show,
alternately, the projections
of the $M_{bc}$ and $\Delta E$ signal bands for
the selected $B\rt\jp K_1(1270)$ candidates 
in the $\jp K^+\pipi$, 
$\jp K^+\pi^-\pi^0$,
and $\jp K^0\pipi$ channels, respectively.  
The $M_{bc}$ distributions are for events 
with $\vert\Delta E\vert <0.06$~GeV,
and the $\Delta E$ distributions are
for events with $M_{bc}>5.271$~GeV.
The curve in each figure is the result of a simultaneous
fit to the $M_{bc}$ and $\Delta E$ projections where
the two distributions are
fitted with Gaussian signal functions 
that are constrained to have the same number of
events.  For the $M_{bc}$ projection, we parameterize the
background with a function that behaves like phase space
near the endpoint;
for $\Delta E$, we represent the background with a 
linear function and restrict the fit to the range 
$-0.1~{\rm GeV} < \Delta E < +0.2$~GeV~\cite{sideband}.
The widths of the $M_{bc}$ distributions are primarily due to the
cm energy spread and are expected to be nearly
the same for each channel; in the fits we fix them at
the value $\sigma_{M_{bc}}=3.0$~MeV, which is 
determined from $B^0\rt\jp K^0_S$ events
in the same data sample~\cite{sin2phi1}. 
For the fits to the $K^+\pipi$ channel, the 
values of the signal peak positions and the $\Delta E$ width are 
free parameters; the fit results are
consistent with MC expectations.  In the $K^+\pi^-\pi^0$ and
$K^0\pipi$ channels, where the statistics are limited, the
signal peak positions and widths are fixed at their expected values.

We use the $B^+ \rt \jp K^+ $ events from the same data sample for 
normalization. 
We select these events using the same
$\jp$ and charged kaon criteria as used in the $B\rt \jp K\pi\pi$
selection.  The number of events are extracted 
using the same fitting procedure.
The yields from the fits 
for all channels are listed in Table~\ref{table:fit_results}.

We searched for a $B^+\rt\jp K^+_1(1400)$ signal using selection
requirements optimized for $K^+_1(1400)\rt K^{*0}\pi^+\rt K^+\pipi$,
namely $\vert M_{K\pi} - M_{K^*}\vert < 50$~MeV and 
$\vert M_{K\pi\pi}-M_{K_1(1400)}\vert <175$~MeV.
In this case the signal observed is consistent with the contribution
from the tail of the $K_1(1270)$.
(There is about a 40\% overlap between the 
$K_1(1270)$ and the $K_1(1400)$ selection
requirements.)  We 
determine an acceptance-corrected ratio of the event yield
$N_{ev}[B^+\rt \jp K^+_1(1400)]/N_{ev}[B^+\rt \jp K^+_1(1270)]
= 0.07 \pm 0.14$ from which
we conclude that contributions to
the $K_1(1270)$ signal from the $K_1(1400)$ resonance
are less than 9\% (at the $1\sigma$ level).

The number of $K^+_1(1270)$ events 
in the $\pipi$ {\it vs} $K^+\pipi$ mass window,
determined from the fit to
the $M_{K\pi\pi}$ distribution 
shown in Fig.~\ref{fig:Mpipi_vs_MKpipi}(b), is $53.2\pm 10.0$ events,
which is very nearly the same as the number of signal events determined
from the simultaneous fits to the $\Delta E$ and $M_b$ projections 
($53.4$ events).  From this agreement, we rule out more than 
a 7\% non-resonant $K\pi\pi$ component to our observed
signal (at the $1\sigma$ level).

We determine the ratio of branching fractions 
using MC-determined  acceptances 
and $K_1$ branching fractions to
the accepted topologies that are taken from the PDG tables~\cite{PDG}.
Here we assume   
the ratio of charged to neutral $B$ meson production at the
$\Upsilon(4S)$ is unity.
The results for the two neutral $K_1^0$ modes are
\begin{eqnarray*}
\frac{{\cal B}(B^0\rt \jp K^0_1(1270))}{{\cal B}(B^+\rt \jp K^+)} 
    & = &  1.42 \pm 0.42~~~~~(K^+\pi^-\pi^0~{\rm mode})\\
\frac{{\cal B}(B^0\rt \jp K^0_1(1270))}{{\cal B}(B^+\rt \jp K^+)} 
    & = &  1.07 \pm 0.45~~~~~(\ks\pipi~{\rm mode}),
\end{eqnarray*}
\noindent
where only statistical errors are shown~\cite{statistical}.
Since the results for the two  modes are
consistent within errors, we combine
the two data sets.  The branching fraction ratios for the $B^0$ and
$B^+$ are:
\begin{eqnarray*}
\frac{{\cal B}(B^0\rt \jp K^0_1(1270))}{{\cal B}(B^+\rt \jp K^+)} 
    & = &  1.30 \pm 0.34 \pm  0.28 \\
\frac{{\cal B}(B^+\rt \jp K^+_1(1270))}{{\cal B}(B^+\rt \jp K^+)} 
    & = &  1.80 \pm 0.34 \pm  0.34,
\end{eqnarray*}
where the first errors are statistical and the 
second are systematic.
The absence of any signal in the $B^+\rt\jp K^+_1(1400)$ channel
translates to a 90\% confidence level limit on the branching fraction
ratio of
\[
\frac{{\cal B}(B^+\rt \jp K^+_1(1400))}{{\cal B}(B^+\rt \jp K^+_1(1270))} 
< 0.30.
\]

The largest component  of the systematic error ($\pm 14\%$)
is due to errors in the $K_1(1270)$ branching fractions 
to the $K\pi\pi$ modes that are
used for this measurement.  We also include 
in the systematic error the level of possible
contributions from other $K\pi\pi$ resonances ($\pm 9\%$)
and non-resonant $K\pi\pi$ production ($\pm 7\%$), uncertainties in
the relative $\jp K\pi\pi$ and $\jp K^+$ acceptance ($\pm 5\%$ for
the $B^+$ channel and $\pm 10\%$ for $B^0$)
and, for the $B^0$, the uncertainty in
the ratio of charged to neutral $B$ meson production at the
$\Upsilon(4S)$ ($\pm 8\%$)~\cite{fplus}.

Since the $B\rt \jp \ks\pipi $ decays 
proceed primarily via the {\em flavor-nonspecific} $\ks\rho^0$ 
intermediate state, they are potentially useful for $CP$ violation 
studies.  In principle, these final states are
mixtures of $CP=\pm 1$ eigenstates, depending on the orbital
angular momentum of the $\jp$ and the $K_1$.
With sufficient statistics, the relative strengths of the
two $CP$ eigenstates can be determined from an analysis
of final state helicity angle distributions~\cite{dunietz}.  
The situation is made complicated by possible interference
between the $\ks\rho$ amplitude and those for
$K^{*}\pi$ and $K_0^{*}(1430)\pi$.  Theoretical work is needed 
to clarify the situation.

In summary, we report the first observation of the
$B\rt\jp K_1(1270)$ decay mode. 
Using the PDG value of
${\cal B}(B^+\rt\jp K^+) = 1.00\pm 0.10 \times 10^{-3}$~\cite{PDG},
we translate our measurements into the branching fractions:
\begin{eqnarray*}
{\cal B}(B^0\rt\jp K^0_1(1270))
    & = &  (1.30 \pm 0.34 \pm  0.31)\times 10^{-3} \\
{\cal B}(B^+\rt\jp K^+_1(1270))
    & = &  (1.80 \pm 0.34 \pm 0.39)\times 10^{-3}.
\end{eqnarray*}
\noindent
These measurements indicate that
this mode constitutes a reasonable portion ($\sim 15\%$) of
the total number of $B\rt\jp X$ decays.  
We see no evidence 
for $B\rt\jp K_1(1400)$ and set an upper limit for this decay
branching fraction.

We wish to thank the KEKB accelerator group 
for the excellent
operation of the KEKB accelerator.
We acknowledge support from the Ministry of Education, Science, Sports and
Culture of Japan and
the Japan Society for the Promotion of Science;
the Australian Research Council and the Australian Department of Industry,
Science and Resources;
the Department of Science and Technology of India;
the BK21 program of the Ministry of Education of Korea, CHEP SRC program
of the
Korea Science and Engineering Foundation and Creative Research Initiative
program of the Ministry of Science and Technology of Korea;
the Polish State Committee for Scientific Research 
under contract No.2P03B 17017; 
the Ministry of Science and Technology of Russian Federation;
the National Science Council and the Ministry of Education of Taiwan;
the Japan-Taiwan Cooperative Program of the Interchange Association;
and the U.S. Department of Energy.

%
%

\begin{table}[tb]
\begin{center}
\caption{Results of the fits to the $M_{bc}$ and $\Delta E$ projections. }
\label{table:fit_results}
\begin{tabular}{lc}
\hline
Channel                        &       $N_{evts}$          \\\hline
$\jp K^+\pipi$                 &       $53.4 \pm 9.1$        \\
$\jp K^+\pi^-\pi^0$            &       $19.3 \pm 5.1$        \\
$\jp \ks\pipi$                 &       $6.2  \pm 2.6$       \\\hline
$\jp K^+$                      &       $472.4 \pm 22.9$      \\\hline
\end{tabular}
\end{center}
\end{table}

%
%


\begin{references}
\bibitem{sideband}
The negative $\Delta E$  region contains events
from $B\rt\jp K \pi\pi X$ where $X$ is a low energy pion or photon
that is not included in the $M_{bc}$ and $\Delta E$ computation.  
Therefore, we do not use this region for sideband
studies
and exclude the $\Delta E <-0.1$~GeV region from the fits shown
in Fig.~\ref{fig:de_mb_fits}.
\bibitem{himass} The range of the fit is restricted to
$M_{K\pi\pi} < 1.7$~GeV
because of possible contributions from higher 
mass $K\pi\pi$ states.
\end{references}
\end{document}